\documentclass[9pt,preprint]{sigplanconf}

\usepackage{amsmath}
\usepackage{amsfonts}
\usepackage{amssymb}
\usepackage{amsthm}
\usepackage{url}
\usepackage{semantic}
\usepackage{stmaryrd}

\newcommand{\eg}{\textit{e.g.}}
\newcommand{\ie}{\textit{i.e.}}
\newcommand{\D}{\mathsf{D}}
\newcommand{\T}{\mathsf{T}}

\newcommand{\interp}[1]{\llbracket{#1}\rrbracket}

\title{The semantic marriage of monads and effects
 \\ {\large{Extended abstract}}}

\authorinfo{Dominic Orchard \and Tomas Petricek \and Alan Mycroft}
           {Computer Laboratory, University of Cambridge}
           {\{firstname.lastname\}@cl.cam.ac.uk}

\newcommand{\op}{\mathsf{op}}

\newcommand{\tr}{\mathbf{t}}
\newcommand{\fa}{\mathbf{f}}

\begin{document}
\maketitle

% Comments from Ohad: 
% A meta-comment. Here is my internal check list when suggesting
% a new semantic structure: 
% (I) Is the new structure well-known in some other community, hopefully
% possesing a well-established associated theory (e.g. monads with Moggi, 
% Param monads for Atkey).
% (II) The new concept is *usable*. How easy is it to establish that
% some structure satisfies the properties/axioms this concept requires
% e.g. monads are very usable. Wadler's indexed monads are *not* 
% (O(2^n) different monads and O(n2^n) monad morphisms).
% Bonus points if your new concept is usable whereas previous work isn't.
% (III) Your concept/axioms subsumes everything that was done before.
% For example, Moggi's account lets him do the meta-theory of various
% effect lambda-calculi in a uniform manner.  So you can do everything
% people do with effect type systems? e.g. optimization, soundness? 
% In other waords, have you got a theorem other than "every other definition
% is an instance"?
% Really, I am skeptical about a new concept until it satisfies *at least*
% one of those properties 

% This differs somewhat to Atkey's approach with a \emph{parameterised
% monad} defined for $T : \mathcal{S} \times \mathcal{S}^\op \rightarrow [\mathcal{C}, \mathcal{C}]$. Our indexed monad structure arises simply by preserving the monoidal
% structure of effect annotations in the monoidal structure of endofunctor
% composition. 

\noindent
\begin{abstract}
  Wadler and Thiemann unified \emph{type-and-effect systems} with
  \emph{monadic semantics} via a syntactic correspondence and
  soundness results with respect to an operational semantics. They
  conjecture that a general, ``coherent'' denotational semantics can
  be given to unify effect systems with a monadic-style semantics. We
  provide such a semantics based on the novel structure of an
  \emph{indexed monad}, which we introduce. We redefine the semantics
  of Moggi's computational $\lambda$-calculus in terms of (strong)
  indexed monads which gives a one-to-one correspondence between
  indices of the denotations and the effect annotations of traditional
  effect systems. Dually, this approach yields \emph{indexed comonads}
  which gives a unified semantics and effect system to
  \emph{contextual} notions of effect (called \emph{coeffects}), which we
  have previously described~\cite{petricek2013coeffects}. \\
\end{abstract} 

\noindent
Previously, Wadler and Thiemann established a syntactic correspondence
between \emph{type-and-effect systems} and the \emph{monadic semantics} 
approach by annotating monadic type constructors with the effect sets of the
type-and-effect system~\cite{wadler2003marriage}. They established
soundness results between the effect system and an operational
semantics, and conjectured a ``coherent semantics'' of effects and
monads in a denotational style. One suggestion was to associate to
each effect set $\sigma$ a different monad $T^\sigma$.

We take a different approach to a coherent semantics, unifying effect
systems with a monadic-style semantics in terms of the
novel notion of \emph{indexed monads}, which generalises monads.\footnote{Note this differs to
  Johnstone's notion of indexed monad in the context of topos theory,
  the indexed monads seen in
  the work of McBride~\cite{mcbridefunctional}, and parameterised
monads by Atkey~\cite{atkey2006parameterised}.}

\paragraph{Indexed monads}

Indexed monads comprise a functor
\[
\T : \mathcal{I} \rightarrow
[\mathcal{C}, \mathcal{C}]
\]
(\ie{}, an indexed family of endofunctors)
where $\mathcal{I}$ is a strict monoidal category $(\mathcal{I}, \otimes, 1)$
and $\T$ is a \emph{lax monoidal functor}, mapping the strict monoidal structure
on $\mathcal{I}$ to the strict monoid of endofunctor composition $([\mathcal{C}, \mathcal{C}], \circ, I_{C})$.

\noindent
The operations of the lax monoidal structure are thus:
\begin{align*} 
\eta_1 : I_\mathcal{C} \xrightarrow{.} \T 1
\quad & \quad
\mu_{F,G} :\T F \circ \T G \xrightarrow{.} \T (F \otimes G)
\end{align*}

\noindent
These lax monoidal operations of $\T$ match the shape of the regular monad operations.
Furthermore, the standard associativity and unitality conditions of the lax monoidal
functor give coherence conditions to $\eta_1$ and $\mu_{F,G}$ which
are analogous to the regular monad laws, but with added
indices, \eg{}, $\mu_{1,G} \circ (\eta_1)_{\T  G} = id_{\T G}$.

\paragraph{Example} (Indexed exponent/reader monad) 
Given the monoid $(\mathcal{P}(X), \cup, \emptyset)$ (for some set $X$), 
 the indexed family of $\mathbf{Set}$ endofunctors 
where $\T X A = X \Rightarrow A$ (with $\Rightarrow$ denoting exponents) and 
$\T X f = \lambda k . f \circ k$, is an indexed monad with:
\begin{align*}
\eta_{\emptyset} a & = \lambda x . a \\
\mu_{F,G} k & = \lambda x . (k \, (x - (G - F))) \, (x - (F - G))
\end{align*}
where $x : F \cup G$ and $k : F \Rightarrow (G \Rightarrow A)$ 
thus $k$ takes two arguments, the $F$-only subset of $x$ (written $x - (G - F)$)
and the $G$-only subset of $x$ (written $x - (F - G)$) where ($-$) is set difference.

The indexed reader monad models the composition of computations with 
implicit parameters, where the required implicit parameters of 
subcomputations are combined in their composition. This 
provides a more refined model to the notion of
implicitly parameterised computations than the traditional reader monad,
where implicit parameters are uniform throughout a computation and its
subcomputations. 

\paragraph{Relating indexed monads and monads}

Indexed monads collapse to regular monads when $\mathcal{I}$ is a
single-object monoidal category. Thus, indexed monads generalise monads.

Note that indexed monads are \emph{not} indexed families of monads. 
That is, for all indices $F \in obj(\mathcal{I})$ then 
$\T F$ may not be a monad.
%However, if $\otimes$ is \emph{idempotent} 
%then indexed monads collapse to indexed families of monads
%since for all $F$, $F \otimes F = F$ and therefore $\T F$ is a monad
%with 
%$\mu_{F,G} :\T F \circ \T F \xrightarrow{.} \T F$ is defined.

\paragraph{An indexed monadic semantics for $\lambda_c$}

We extend indexed monads to \emph{strong} indexed monads, with an 
indexed strength operation (and analogous laws to usual monadic
strength):
\begin{align*}
(\tau_{F})_{A,B} : (A \times \T F B) \rightarrow \T F (A \times B)
\end{align*}
\noindent
We replay Moggi's categorical semantics for the computational 
$\lambda$-calculus ($\lambda_c$)~\cite{moggi1989monads}, replacing the regular
strong monad operations with the analogous operations of an indexed strong monad.
This provides an indexed semantics. For example, the semantics
of $\lambda$-abstraction becomes the following (where we write the parameter to 
$\T$ as a subscript for notational clarity below):
\begin{align*}
\frac
{\interp{\Gamma, x : \sigma \vdash e : \tau} = 
g : \interp{\Gamma} \times \interp{\sigma} \rightarrow {\T_F} \, \tau}
{\interp{\Gamma \vdash \lambda x : \sigma . e : \sigma \rightarrow \tau} 
= \eta_1 \circ (\Lambda g) : \interp{\Gamma} \rightarrow \T_1 \,
(\sigma \Rightarrow \T_F \, \tau)} 
\end{align*}
(where for $g : A \times B \rightarrow C$, $\Lambda g : A \rightarrow (B \Rightarrow C)$).
\noindent

\paragraph{Coherent semantics}
In this indexed monadic semantics, the indices of denotations have
exactly the same structure as the effect annotations of a traditional
effect system (with judgments $\Gamma \vdash e : \tau, F$ for an
expression $e$ with effects $F$).

We unify effect systems with indexed monadic semantics, so that
$\interp{\Gamma \vdash e : \tau, F} : \interp{\Gamma} \rightarrow \T_F
\, \interp{\tau}$, taking $obj(\mathcal{I})$ as the effect sets of a
traditional effect system, with the strict monoidal structure on
$\mathcal{I}$ provided by the effect lattice, with $1 = \bot$ and
$\otimes = \sqcup$, and morphisms $f : X \rightarrow Y$ in
$\mathcal{I}$ iff $X \sqsubseteq Y$ in the effect lattice.
Pleasingly, the usual equational theory for $\lambda_c$ (such
as $\beta$-equality for values) follows directly from the 
strong indexed monad axioms.
  %Thus, $\interp{\Gamma \vdash e :
%  \tau, F} : \interp{\Gamma} \rightarrow \T_F \, \interp{\tau}$.

The morphism mapping of $\T$ defines natural transformations
$\iota_{X,Y} : \T X \xrightarrow{.} \T Y$ when $X \sqsubseteq
Y$ which provides a semantics to sub-effecting:
\begin{equation*}
\inference[(sub)]{\interp{\Gamma \vdash e : \tau, F'} = g : \interp{\Gamma} \rightarrow \T_{F'} \interp{\tau} & F' \sqsubseteq F}
          {\interp{\Gamma \vdash e : \tau, F} = \iota_{F',F} \circ g :  \interp{\Gamma} \rightarrow \T_F \interp{\tau}}
\end{equation*}
For a particular notion of effect, the indexed strong monad can be
defined such that the propagation of effect annotations in an effect
system maps directly to the semantic propagation of effects. 
For example, for memory effects the functor can be made more \emph{precise}
with respect to the effect, \eg{}, $\T \{\mathsf{read} \, \rho : \tau\}
\, A = \tau \rightarrow A$ and $\T \{\mathsf{write} \, \rho : \tau\}
\, A = A \times \tau$ (note: the latter is not itself a monad). 

Therefore strong indexed monads neatly unify a (categorical) semantics of effects
with traditional effect systems. The indexed monad structure arises
simply from the standard category theory construction of lax monoidal functors, where $\T$
preserves the strict monoidal structure of $\mathcal{I}$ in
$[\mathcal{C}, \mathcal{C}]$.  Crucially, indexed monads are
\emph{not} an indexed family of monads (contrasting with Wadler and
Thiemann's original conjecture).%, that is, for all $X$, the endofunctor $\T \, X
%: \mathcal{C} \rightarrow \mathcal{C}$ may not be a monad.

\paragraph{In context}

We argue our approach provides an intermediate solution between the
traditional monadic approach (which does not couple annotations of
an effect system to semantics) and algebraic effect
theories (see, \eg{}, Kammar and Plotkin~\cite{kammar2012algebraic}).

%We connect the indexed monad/joinad approach to more recent
%work with program logics, such as Nanevski \emph{et al.}'s work on
%\emph{Hoare Type Theory} where monads are effectively annotated with
%pre- and post-condition pairs~\cite{types-hoare-theory}.

Our approach differs somewhat to Atkey's \emph{parameterised monads},
defined for $T : \mathcal{S} \times \mathcal{S}^\op \rightarrow
[\mathcal{C}, \mathcal{C}]$. Our indexed monad structure has a more
systematic derivation, arising from the strict monoidal preservation of the 
lax monoidal functor. This technique can be applied to derive
coherent semantic structures/effect system pairs for other notions of
computation.

Effect systems traditionally define effect annotations in terms 
of sets with composition via set union~\cite{gifford1986effects}. This has
the additional property that combining effect annotations is symmetric (due
to commutativity of union). The more general structure of a monoid here,
also used by Nielson and Nielson~\cite{nielson1999type}, provides an opportunity
for generating effect information that records the \emph{order} of effects.

\paragraph{Extending the approach to other notions}

We apply the same technique used to derive indexed monads 
to give richer effect systems/semantics in two ways.

\begin{enumerate}
\item{A strict \emph{colax} monoidal functor $\D :
\mathcal{I} \rightarrow [\mathcal{C}, \mathcal{C}]$ gives rise to the
dual notion of \emph{indexed comonads}, which we have previously shown
to provide the notion of a \emph{coeffect} system (analysing contextual
requirements) and a semantics for contextual
program effects~\cite{petricek2013coeffects,orchard2013thesis}.

For a monoid $(\mathcal{I}, \otimes, 1)$, 
indexed comonads have the colax operations:
\begin{align*} 
\varepsilon_1 : \D 1 \xrightarrow{.} I_\mathcal{C}
\quad & \quad
\delta_{F,G} :\D (F \otimes G) \xrightarrow{.} \D F \circ \D G 
\end{align*}
Interestingly, indexed comonads seem much more useful than comonads
since they relax the usual \emph{shape preservation} property of comonads.

\paragraph{Example} (Indexed partiality comonad)
For the boolean conjunction monoid $(\{\fa, \tr\}, \wedge, \tr)$, 
the following indexed family of endofunctors is an
indexed comonad:
\begin{align*}
\begin{array}{lll}
& \D\ \fa A = 1 & \quad \D\ \tr A = A \\
& \D\ \fa f =\ !_A & \quad \D\ \tr f = f
\end{array}
\end{align*}
with $\varepsilon_{\tr} a = a$, $
\delta_{\tr, \tr} a = a$, and $\delta_{X,Y} a = 1$ when $X = \fa$ and/or
$Y = \fa$. The indexed partiality comonad is essentially a
dependently-typed partiality construction $\D A = 1 + A$. Note
however, that $\D A = 1 + A$ is not a comonad since $\varepsilon : \D
A \rightarrow A$ is not well defined.  The indexed partiality comonad
encodes the notion of a partial context/input to a computation, and
has been previously shown to give a simple liveness analysis coupled 
with a semantics that embeds the notion of 
dead-code elimination~\cite{petricek2013coeffects}.

Coeffect systems differ to effect systems in their 
treatment of $\lambda$-abstraction where, for coeffect judgments 
$\Gamma ? F \vdash e : \tau$ (meaning expression $e$ has contextual requirements
$F$):
\begin{equation*}
\inference[(abs)]{\Gamma, x : \sigma ? F \vee F' \vdash e : \tau}
          {\Gamma ? F \vdash \lambda x . e : \sigma \xrightarrow{F'} \tau}
\end{equation*}
Reading this rule top-down, the coeffects of the function body are split 
between \emph{immediate} contextual requirements and \emph{latent} requirements
(written $\xrightarrow{F'}$).
Thus, with respect to contextual requirements, $\lambda$-abstraction
is not ``pure'' as it is for effects.
In an indexed semantics unifying a coeffect
system with an indexed comonad, the semantics of $\lambda$-abstraction requires the additional structure
of an \emph{indexed (semi-)monoidal comonad} with the operation:
\begin{equation*}
(m_{F,G})_{A,B} : \D_F A \times \D_G B \rightarrow \D_{F \vee G} (A \times B)
\end{equation*}
where $\vee$ is an associative binary operation over $\mathcal{I}$.
}
\item{Nielson and Nielson defined a more general effect system with a
richer algebraic effect structure, separating the traditional approach
of an effect lattice into operations for sequential composition,
alternation, and fixed-points~\cite{nielson1999type}. Relatedly on the semantic
side, the structure of a \emph{joinad} has been proposed to give the
semantics of sequencing, alternation and parallelism in an effectful
language~\cite{joinads-haskell11}, adding additional monoidal
structures to a monad. Similarly to indexed monads, 
joinads can be generalised to \emph{indexed joinads}, giving a correspondence between
the richer effect systems of Nielson and Nielson and a joinad-based 
semantics. This is future work.}
\end{enumerate}

\bibliography{references}

\end{document}